\documentclass[aps,pre,floats,preprint,showpacs,superscriptaddress]{revtex4}

\usepackage{graphicx,epsfig}
\usepackage{times}
\usepackage{graphics,dcolumn,bm,fleqn,float}
\usepackage{amssymb,amsmath,multirow,rotate,color}
\begin{document}

\title{Robustness of Cooperation in the Evolutionary Prisoner's
  Dilemma on Complex Networks}

\author{J. Poncela}

\affiliation{Institute for Biocomputation and Physics of Complex
Systems (BIFI), University of Zaragoza, Zaragoza 50009, Spain}

\author{J. G\'{o}mez-Garde\~{n}es}

\affiliation{Institute for Biocomputation and Physics of Complex
Systems (BIFI), University of Zaragoza, Zaragoza 50009, Spain}

\affiliation{Departamento de F\'{\i}sica de la Materia Condensada,
University of Zaragoza, Zaragoza E-50009, Spain}

\author{L. M. Flor\'{\i}a}

\affiliation{Institute for Biocomputation and Physics of Complex
Systems (BIFI), University of Zaragoza, Zaragoza 50009, Spain}

\affiliation{Departamento de F\'{\i}sica de la Materia Condensada,
University of Zaragoza, Zaragoza E-50009, Spain}

\author{Y. Moreno}

\email{yamir@unizar.es}

\affiliation{Institute for Biocomputation and Physics of Complex
Systems (BIFI), University of Zaragoza, Zaragoza 50009, Spain}

\date{\today}

\begin{abstract}

Recent studies on the evolutionary dynamics of the Prisoner's
Dilemma game in scale-free networks have demonstrated that the
heterogeneity of the network interconnections enhances the
evolutionary success of cooperation. In this paper we address the
issue of how the characterization of the asymptotic states of the
evolutionary dynamics depends on the initial concentration of
cooperators. We find that the measure and the connectedness
properties of the set of nodes where cooperation reaches fixation
is largely independent of initial conditions, in contrast with the
behavior of both the set of nodes where defection is fixed, and
the fluctuating nodes. We also check for the robustness of these
results when varying the degree heterogeneity along a
one-parametric family of networks interpolating between the class
of Erd\"{o}s-Renyi graphs and the Barab\'asi-Albert networks.

\end{abstract}


\maketitle

\section{Introduction}
\label{intro}

Evolutionary dynamics has proved to be a useful theory to describe
evolution of biological systems at all levels of organization
\cite{nowak_book}. Rooted in the basic tenet of Darwinism, the
replicator dynamics \cite{hofbauer,gintis,sigmund} of evolutionary
game theory provides an elegant mathematical description of how
natural selection among (phenotypes) strategies takes place when
the reproductive success of individuals (and then the future
abundance, {\em i.e.} frequency, of strategies) depends on the
current phenotypic composition of the population
(frequency-dependent fitness). In this regard, one of the current
theoretical challenges to the explanatory powers of evolutionary
game dynamics is the understanding of the observed evolutionary
survival of cooperative behavior among individuals when selfish
actions provide a higher benefit (fitness). Perhaps the best
suited (and most used) model to formally describe the puzzle of
how cooperation arises is the Prisoner's Dilemma (PD), a
two-players-two-strategies game, where each player chooses one of
the two available strategies, cooperation or defection: A
cooperator receives $R$ when playing with a cooperator, and $S$
when playing with a defector, while a defector earns $P$ when
playing with a defector, and $T$ against a cooperator. When
$T>R>P>S$, the game is a PD. Given this payoff ordering, in a
well-mixed (unstructured) population where each agent interacts
with all other agents (or a representative sample of the
population composition), defectors are fitter and thus the
fraction of cooperators asymptotically vanishes.

Among the various mechanisms that have been proposed to explain
how natural selection can lead to cooperative behavior (like kin
selection, group selection, direct or indirect reciprocity)
\cite{nowak06}, a simple one is based on leaving off the
well-mixed population hypothesis, so that each individual only
interacts with agents in its neighborhood, as specified by some
graph or network of "social" interactions. Agent-based-modelling
approaches \cite{axelrod} of this kind in Theoretical Biology
\cite{Ecology}, Economics \cite{Judd} and Social Sciences
\cite{Schelling} often benefit in a natural way from Statistical
Physics methods, concepts and techniques (also scientists), so
favoring fruitful (synergic) interdisciplinary (socio-, bio-,
econo-) physics research \cite{szabo}, often termed Physics of
Complex Systems \cite{anderson,boccara}.

Early pioneering numerical work \cite{nowak} on the PD game in
two-dimensional square lattices, made the observation that, unlike in
unstructured populations, cooperators and defectors can coexist in the
lattice indefinitely. In \cite{nowak} each individual node played with
its immediate neighbors each time step accumulating a payoff, then
updated its strategy by imitating the one of highest payoff in its
neighborhood, including itself (best-takes-over reproduction rule) and
back again for very large times. When passing from a "mean field"
(well mixed population) interaction description to a lattice structure
of interactions, one has to specify various details (of varying
importance) on both, {\em a)} the lattice characteristics, {\em e.g.}
regular or not, randomness of various kinds, finite size effects,
etc., and {\em b)} the specific form of the microscopic dynamics of
reproduction process, {\em e.g.} deterministic rules or probabilistic
ones, synchronous or asynchronous updating, what types of stochastic
fluctuations are allowed, etc. The study of many, if not most, of
important aspects of the issue have generated for more than a decade a
wealthy literature, of a great interest from the Statistical Physics
perspective ({\em e.g.},
\cite{huberman,nowakbo,lindgren,nakamaru,toke,abramson,sh,hs,ifti,doebeli,duran,hs05,svs,vss,perc};
for a recent review, see \cite{szabo}, where an extensive list of
references can be found).

Nowadays, the existence of {\em cooperation-promoting feedback
mechanisms that are rooted deep into the interaction structure} is
indisputably accepted. It has been termed {\em spatial}, or {\em
lattice} reciprocity, in analogy to {\em direct} reciprocity
(through iterated game strategies) and {\em indirect} reciprocity
(through reputation, or scoring, of agents). Simply said, the
clustering of cooperators in the lattice could provide high enough
fitness to the cooperator nodes exposed to invasion, to the extent
of preserving cooperators from evolutionary extinction, even when
defection is blatantly favored by the one-shot (two-players) game
analysis. For negligible values of $P-S\simeq0$, when $T-R$
increases from zero cooperation decreases slowly, and becomes zero
at values of $(T/R)-1$ well beyond zero. The region (in parameter
space) of coexistence of strategists is the genuine battlefield
where the competition between strategies adopts interesting,
non-trivial aspects: The transition region between two clear-cut
phases, {\em i.e.} all-cooperators (all-C) prevailing at
$T/R\simeq1 $, and all-defectors (all-D) at higher values of
$T/R$. More recently, a set of works have extended this
perspective to a most intriguing and ubiquitous class of networks,
say scale-free networks, a "focuss issue" nowadays.

There is an accumulated evidence that many real biological
\cite{JeongNat01,SolePRSLB01}, social \cite{NewmanPNAS01} and
technological \cite{FaloutsosCCR99,vespignanibook,WangPRE06}
systems are neither regular nor simplest random graphs (not to say
well-mixed populations) of entities or agents, but they are
described by some distinctive metric (path length based) and
topological (structure and size of local neighborhoods)
properties. They often show a so-called scale-free (SF)
distribution density of degree, $P(k)\sim k^{-\gamma}$, where the
degree $k$ of a node is the number of connections it shares with
its neighbors \cite{newmanrev,yamirrep}, so their connectivity
patterns depart considerably from lattice homogeneity (lacking of
a sharp characteristic scale of connectivity). The ubiquity and
importance of complex networks raised quite naturally the question
of how natural selection works on top of different types of
complex networks of agents
\cite{abramson,smallworld,maxi,pacheco,pacheco2,prlnoi}. In this
case (as in other nonlinear dynamical processes in networks
\cite{synchronization,michaelis-menten}) one has to deal with two
sources of complexity, the evolutionary dynamics and the complex
structure of the substrate, which are entangled. Interestingly,
the sort of processes that evolutionary game dynamics is aimed to
model may well be very relevant to understand real networked
systems through the study of a variety of scenarios of
co-evolution of both strategies (phenotype survival) and network
(evolving topological features)\cite{maxi}. Among other works
exploring various aspects on the evolution in complex networks,
see \cite{roca,chen,lozano}. From here we focuss attention on
fixed network settings and how degree heterogeneity influences
evolutionary dynamics of PD.

Some recent extensive numerical works on PD (and closely related)
games \cite{pacheco,pacheco2,prlnoi} on SF networks, using
probabilistic updating rule (random neighbor pair-comparison, and
update with probability proportional to fitness difference) have
shown that the absence of a sharp characteristic scale of degree
in the network greatly enhances the "lattice reciprocity"
mechanisms of evolutionary survival of cooperation. For example,
highly connected (hubs) cooperator nodes have the chance of high
payoffs and resist well invasion by easily invading less connected
neighbors, which in turn increase hub's payoffs and invading
capabilities \cite{pacheco2}; this positive feedback mechanism
does not operate in the case of defector hubs and illustrates in a
simple way one of the biasing effects of graph heterogeneity.

In a recent exploration of these heterogeneity based
cooperation-promoting mechanisms, using the kind of implementation
of replicator dynamics on graphs specified above in the previous
paragraph, one observes generically \cite{prlnoi} that fixation of
the cooperation (as well as defection) strategy on certain nodes
occurs after (often-not-large) sensible transients, so that any
asymptotic trajectory of population states defines a partition of
the network into three sets: the set ${\mathcal{C}}$ of nodes
where cooperation is fixed, the set ${\mathcal{D}}$ of nodes where
defection is fixed, and the set ${\mathcal{F}}$ of fluctuating
nodes that experience forever cycles of invasion by the competing
strategies. In other words, the observed stationary value of the
average fraction $\bar{c}$ of cooperators (see definition in
section 3), in any asymptotic (long-term) trajectory, has two
additive contributions: a) the relative size $\mu({\mathcal{C}})$
of the set of pure cooperators, and b) the overall fraction of
time $\bar{T}_c$ spent by fluctuating nodes as cooperators,
weighted by its relative size $\mu({\mathcal{F}})$, say

\begin{equation}
\bar{c} =  \mu({\mathcal{C}}) + \mu({\mathcal{F}}) \bar{T}_c
\label{tachan}
\end{equation}

The analysis of global connectedness inside the sets
${\mathcal{C}}$ and ${\mathcal{D}}$ of fixed strategy nodes
reveals that the lack of a significant characteristic scale of
degree is neatly associated to a simply connected ${\mathcal{C}}$
set, while ${\mathcal{D}}$ is fragmented into many clusters in the
wide transition region (coexistence of strategies) between
asymptotic uniform ($\mu({\mathcal{C}})=1$, all-C, and
$\mu({\mathcal{D}})=1$, all-D) equilibria. This structure of
${\mathcal{D}}$ in the strategies coexistence regime is similar to
that exhibited by both ${\mathcal{C}}$ and ${\mathcal{D}}$ sets
for the Erd\"os-Renyi random class of networks ({\em i.e.} with
Poissonian distribution density of degrees, and thus a significant
characteristic scale: the network average degree) \cite{prlnoi}.
All previous results \cite{prlnoi}, were obtained for a unbiased
($50\%$) initial proportion of (randomly placed) cooperators, for
all the analyzed stochastic trajectories.

In this paper, we are interested in exploring the robustness of these
observations reported in \cite{prlnoi} on the behavior of the
partition sets, for a limiting one-parameter form of the PD game, say
$P-S=0$: the border with the Snowdrift game (see next
section). Robustness against parameter $P-S$ variation, and others,
will be analyzed elsewhere \cite{pnas}. In particular, we focus here
on two aspects of robustness: First, the influence of varying initial
fraction of cooperators on the network partition sets
(${\mathcal{C}}$, ${\mathcal{D}}$, ${\mathcal{F}}$) of asymptotic
trajectories. The model, its dynamical rules and structural
characteristics, as well as the necessary technical details, are the
contents of section \ref{model}. The results are described and
analyzed in section \ref{initial}. Second, in section
\ref{heterogeneity}, we show how those observed behaviors of the
partition vary along an interpolating family of networks whose
heterogeneity can be one-parametric tuned, from ER limit to BA limit,
that is, we check robustness against decreasing heterogeneity of the
network. Conclusions and some prospective remarks can be found in
ending section \ref{conclusions}.

\section{The Model}
\label{model}

The Prisoner's Dilemma game is defined in its more general form by
the payoff matrix:
\begin{equation}
\left( \begin{array}{cc} R & S \\ T & P
\end{array} \right) \label{PD}
\end{equation}
where the element $a_{ij}$ is the payoff received by an
$i$-strategist when playing against a $j$-strategist, with $i=1$
meaning cooperator, and $i=2$ defector. The payoff ordering is
given by $T>R>P>S$. Other payoff orderings have received other
names, {\em e.g.} $T>R>S>P$ corresponds to the so-called Snowdrift
(or Hawks and Doves, or Chicken) game. Following several studies
\cite{nowak,pacheco}, the PD payoffs have been set to $R=1$ (so
the reward for cooperating fixes the payoff scale), $T=b>1$, $P=0$
(no benefit under mutual defection), and $P-S=\epsilon=0$. This
last choice places us in the very frontier of PD game. It has the
effect of not favoring any strategy when playing against defectors
(while being advantageous to play defection against cooperators).
Small positive values of the parameter $\epsilon\ll 1$ leads to no
qualitative differences in the results \cite{nowak,pacheco,pnas},
so the limit $\epsilon \rightarrow 0^+$ is agreed to be
continuous.

The dynamic rule is specified as follows: each time step is
thought of as one generation of the discrete evolutionary time,
where every node $i$ of the system plays with its nearest
neighbors and accumulates the payoffs obtained during the round,
say $P_i$. Then, individuals are allowed to synchronously change
their strategies by comparing the payoffs they accumulated in the
previous generation with that of a neighbor $j$ chosen at random.
If $P_i>P_j$, player $i$ keeps the same strategy for the next time
step, when it will play again with all of its neighborhood. On the
contrary, whenever $P_j > P_i$, $i$ adopts the strategy of $j$
with probability $\Pi_{i\rightarrow j}=\beta (P_j-P_i)$, where
$\beta^{-1}=\text{max}\{k_i,k_j\}b$. Note that this dynamic rule,
though stochastic, does not allow the adoption of irrational
strategy, {\em i.e.} $\Pi_{i\rightarrow j}=0$ whenever $P_j \leq
P_i$.

Let us now specify precisely the family of networks on top of
which the evolutionary PD game is evolved. Strategists are located
on the vertices of a fixed graph of average connectivity $\langle
k \rangle=4$. The heterogeneity of the networks is controlled by
tuning a single parameter $\alpha$, so that when $\alpha=0$ the
networks are of the Erd\"os-Renyi class of random graphs, and when
$\alpha=1$ they are of the Barab\'asi-Albert (BA) \cite{bara}
scale-free networks class. Let's first describe the algorithm to
construct a BA network of size $N$. In this case, one starts from
a fully connected set of $m_0$ nodes and at each time step a new
node is linked to $m=2$ nodes preferentially chosen, namely, the
probability that node $i$ receives one new link is proportional to
its degree, $k_i/\sum_j k_j$. Avoiding multiple connections and
iterating the preferential attachment rule $N-m_0$ times a SF
network with an exponent $\gamma=3$ is generated. On the other
hand, random single-scale networks are built up following the
standard recipe to generate ER networks \cite{yamirrep}. Finally,
networks with an intermediate degree of heterogeneity can be built
following the recipe introduced in \cite{jesus}. The algorithm
combines the mechanisms of preferential (with probability
$\alpha$) and uniform random linking ($1-\alpha$) in such a way
that starting from $\alpha=0$ and increasing its value, the
networks generated are successively more homogeneous with a heavy
tail whose exponent is equal to ($\alpha=0$) or larger than
($\alpha>0$) $\gamma=3$.

From any initial condition $\{s_i(t=0)\}, \;\; i=1,..., N$ (where
$s_i=1$ if node $i$ is an instantaneous cooperator and $s_i=0$ if
defector), and after many generations, the instantaneous fraction
of cooperators $c(t)=N^{-1}\sum_i s_i(t)$ in the stochastic
trajectory, $\{s_i(t)\}$, fluctuates around a well-defined mean
value $\bar{c}$, which depends on the parameter $b$, as well as on
the particular initial condition. The transient time $t_0$ that we
allow before measuring observable quantities is assured to be
larger than the one required for the stationarity of $\bar{c}$
(see below). The average level of cooperation $\langle c \rangle$
is computed as the average of $\bar{c}$ over initial conditions
(of fixed fraction $\rho_0$ of cooperators), and network
realizations. We numerically identify as pure cooperators all
those individuals that {\em always} cooperate, for all times
larger than the transient time $t_0$. Pure defectors are those
that {\em always} defect for any $t>t_0$. Fluctuating nodes are
those that are neither pure cooperators nor pure defectors. In
this way we estimate the measure of the partition sets
(${\mathcal{C}}$, ${\mathcal{D}}$, ${\mathcal{F}}$). To inspect
the connectedness of the sets of pure strategists, ${\mathcal{C}}$
and ${\mathcal{D}}$, we define cooperator ($CC$) and defector
cores ($DC$) as clusters (connected subgraphs) fully composed by
pure cooperators and defectors, respectively, their numbers being
denoted by $N_{cc}$ and $N_{dc}$. It is easy to realize that for
generic (irrational) $b$ values, no pure defector can be a
neighbor of a pure cooperator, so that the presence of both types
of nodes in the long-term stochastic trajectory, assures the
existence of fluctuating nodes.

The time scale of microscopic invasion processes (updating rule)
is controlled by $\beta^{-1}$, which is the highest connectivity
of pair's nodes; this makes that very high payoff of a hub due to
its very high $k$ is sensibly balanced by $\beta\propto k^{-1}$
\cite{pacheco}, with the side effect that the invasion processes
from and to hubs are slowed down, if hub's (and neighbor's) payoff
is much smaller than its connectivity $k$. On the other hand, the
transient time $t_0$ should be greater than characteristic
fixation times for nodes in ${\mathcal{C}}$ and ${\mathcal{D}}$,
if one is interested in measuring observable quantities associated
to the partition. Fixation times of strategies at the nodes in
turn, obviously depends on the initial conditions (i.e., on
$\rho_0$, the initial fraction of cooperators), so that henceforth
in the simulations we use a variable time window, $t_0$, of at
least $10^4$ generations as the transient time. Once the system is
at (a fluctuating) equilibrium regarding stationarity of $\langle
c \rangle$, we let the dynamics evolve for $10^4$ additional time
steps, while measuring quantities. All the results have been
averaged over at least $10^3$ different realizations of the
networks and initial conditions. Most of the results shown below
correspond to $N=4000$ nodes, though other values have been also
used; we will comment on this issue in the concluding section.

\begin{figure}
\begin{center}
\epsfig{file=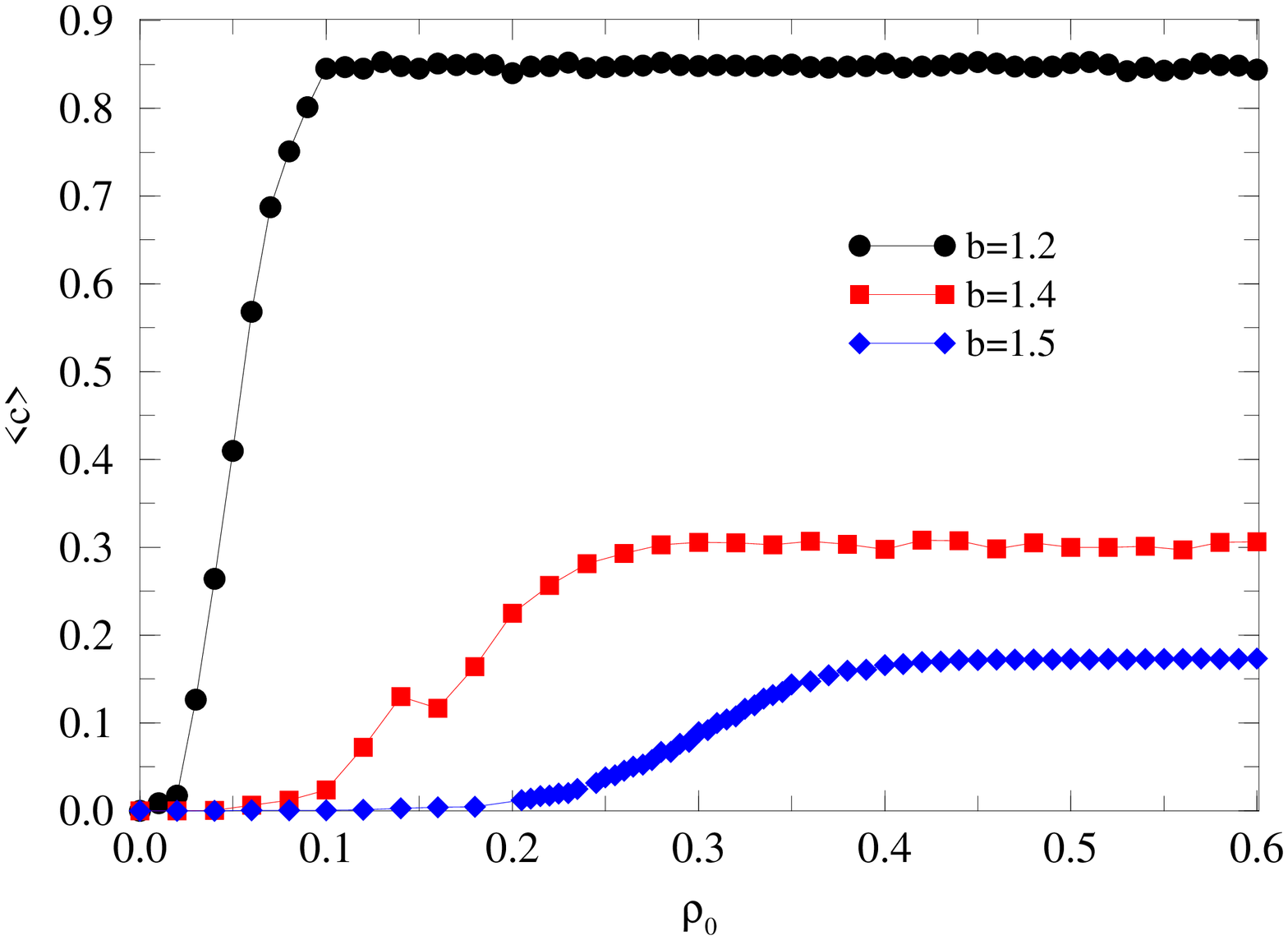,width=2.5in,angle=-90}
\epsfig{file=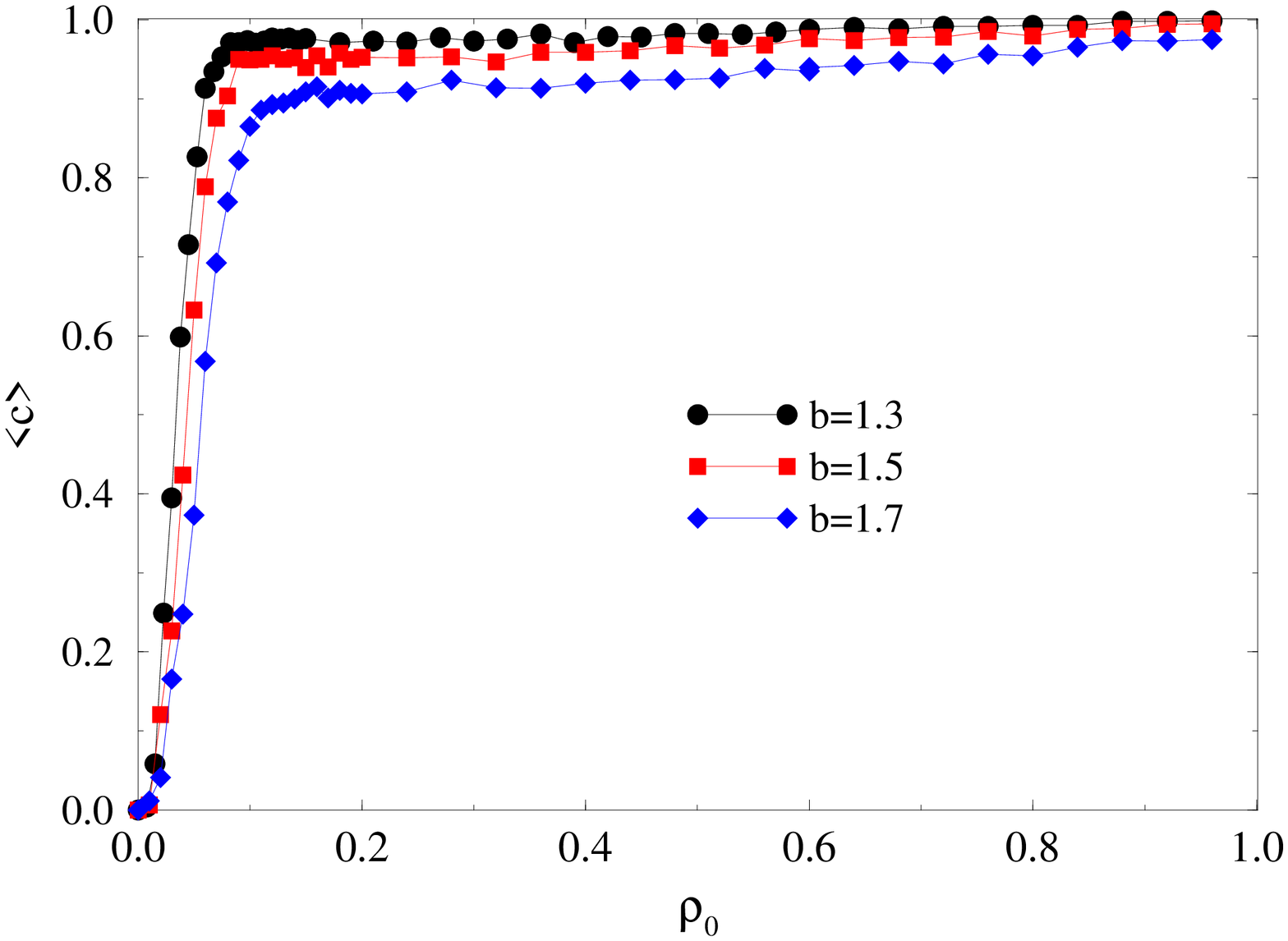,width=2.5in,angle=-90}
\end{center}
\caption{Average cooperation level in ER networks (top panel) and
BA networks (bottom panel) as a function of the initial
concentration $\rho_o$ and several values of $b$ as indicated. The
size of the networks is $N=4000$ nodes and $\langle k \rangle=4$.}
\label{fig2}
\end{figure}

\section{Dependence on the initial conditions}
\label{initial}

The initial conditions for the stochastic trajectories that we
consider here are such that an initial number $\rho_{0}N$ of nodes
($0<\rho_{0}<1$) are randomly chosen as cooperators. In Fig.\
\ref{fig2} we show, for some values of the parameter $b$, how the
stationary value of $\langle c \rangle$ depends on the initial
fraction $\rho_{0}$ of cooperators for ER and BA networks. As seen
in that figure, $\langle c \rangle$ typically increases with
$\rho_{0}$ until saturation is reached much before $\rho_{0}$
approaches $1$. One observes that saturation occurs sooner for
smaller values of $b$. These features are common for both classes
of networks. However some details of the $\langle c
\rangle(\rho_{0})$ curves are different: First, for ER networks,
the departure from zero of $\langle c \rangle (\rho_{0})$ occurs,
as $b$ increases, only above some ($b$-dependent) threshold value
of the initial fraction of cooperators; on the contrary, for BA
networks $\langle c \rangle$ departs from zero as soon as
$\rho_{0}>0$, at all values of $b$ inside the coexistence region.
Second, saturation is more perfect for ER networks, while for BA
graphs the plateau in the $\langle c \rangle(\rho_{0})$ curve has
some small positive slope.

\begin{figure}
\begin{center}
\epsfig{file=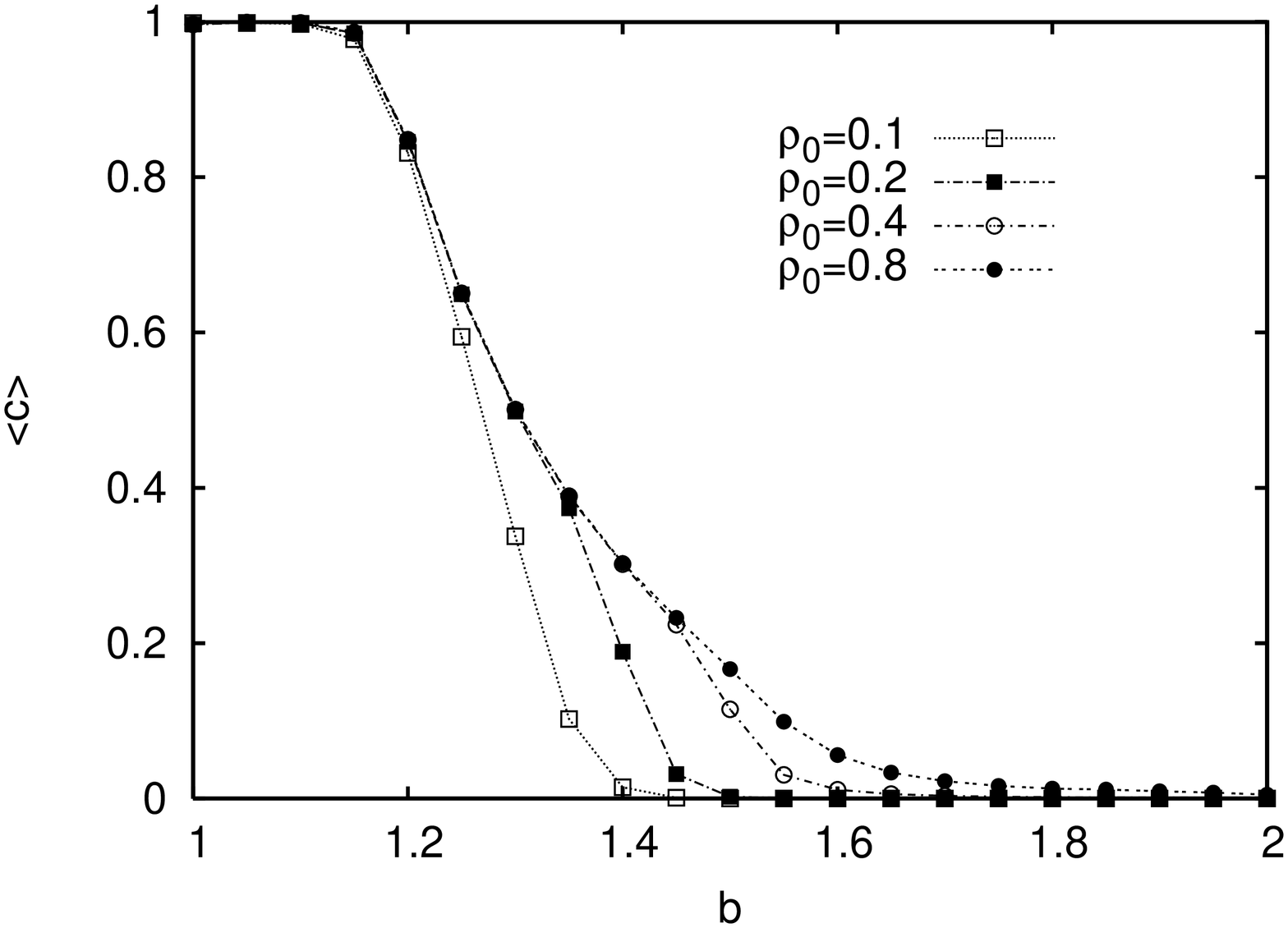,width=2.5in,angle=-90}
\epsfig{file=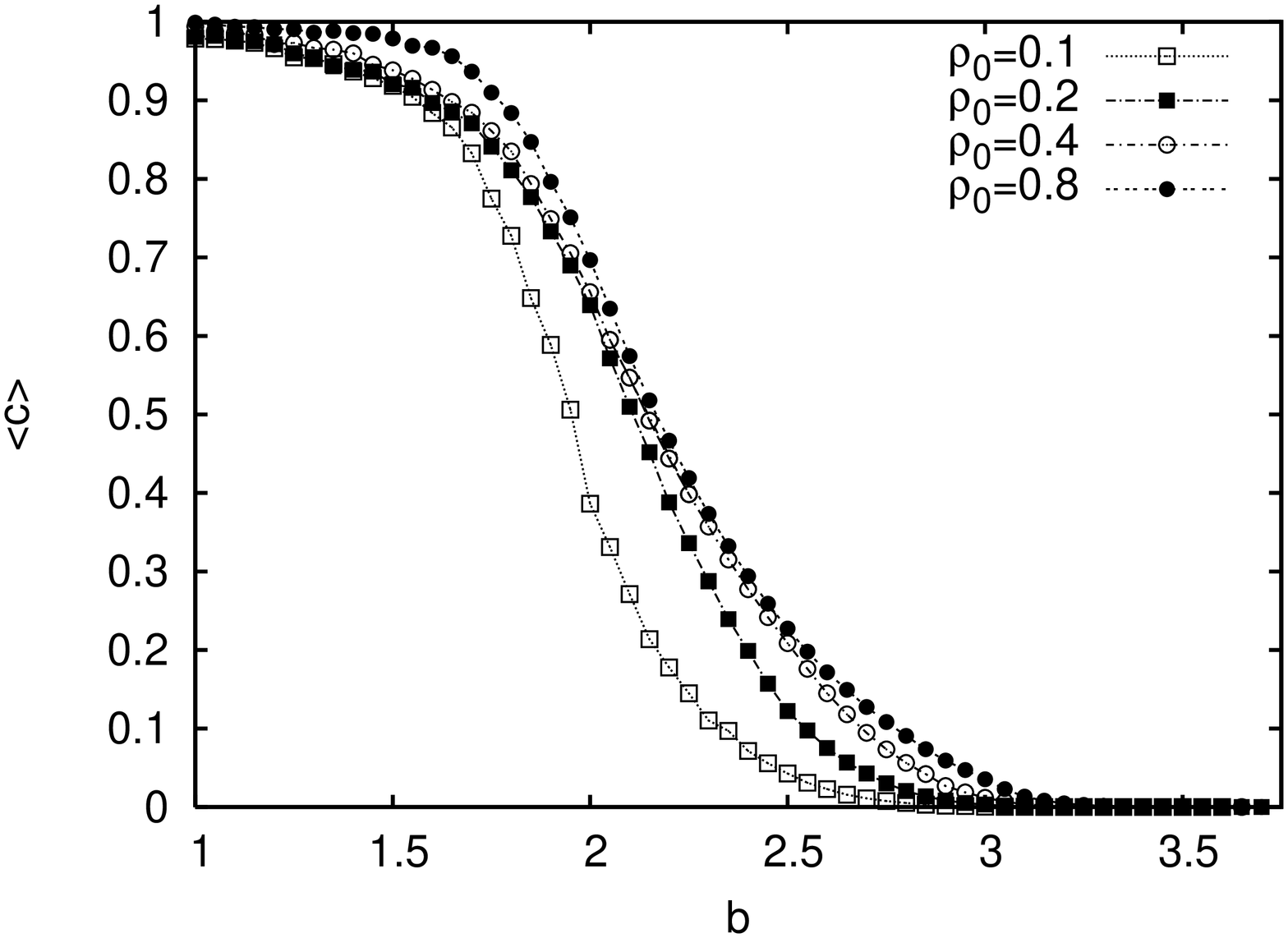,width=2.5in,angle=-90}
\end{center}
\caption{Average cooperation level in ER (top) and SF (bottom)
networks as a
  function of $b$ and different initial concentration of cooperators $\rho_0$
  as indicated. The size of the networks is $N=4000$ nodes and $\langle k
  \rangle=4$. The scale-free network is a BA graph whose $P(k)\sim k^{-3}$.}
\label{fig1}
\end{figure}

The variation with the game parameter $b$ of the stationary
(asymptotic) average cooperation, $\langle c \rangle (b)$, for
several values of $\rho_0$, is shown in Fig.\ \ref{fig1} for ER
graphs (upper panel) and BA networks (lower panel). In the case of
ER networks, different initial concentrations $\rho_0$ produce a
family of curves that mainly differs in their tails in such a way
that the larger the value of $\rho_0$, the slower the decay of
$\langle c \rangle$ as $b$ increases, in correspondence with the
perfect saturation of $\langle c \rangle(\rho_{0})$ at fixed $b$
observed in Fig.\ \ref{fig2}. On the other hand, in BA networks
the effects of different initial conditions are appreciated in the
whole range of $b$ values. We thus see that degree heterogeneity
not only favors the survival of cooperation, but also makes the
value of the average cooperation, at fixed $b$ value, more
dependent on initial conditions. In this regard one should note
that ER networks, often termed as homogeneous, have indeed some
small heterogeneity, {\em i.e.} the degree distribution density
has a non-zero variance. In fact, the average level of cooperation
in ER networks is clearly enhanced with respect to random regular
networks (where all the nodes have exactly the same degree $k$),
see {\em e.g.} \cite{duran}. In other words, even the small
amounts of heterogeneity of ER networks, are enough to allow for
cooperation-promoting feedback mechanisms to work.

\begin{figure}[!tb]
\begin{center}
\epsfig{file=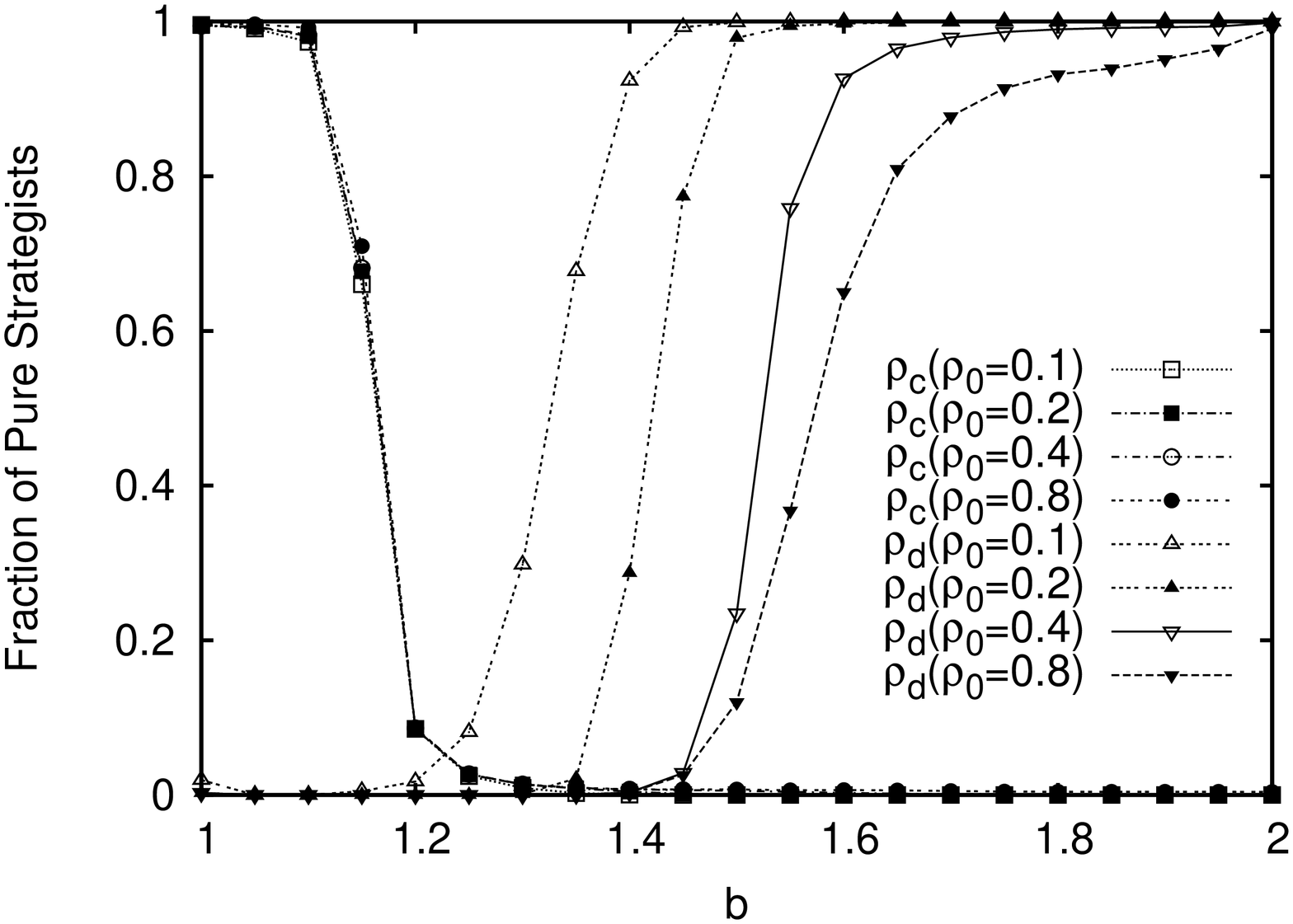,width=2.5in,angle=-90}
\epsfig{file=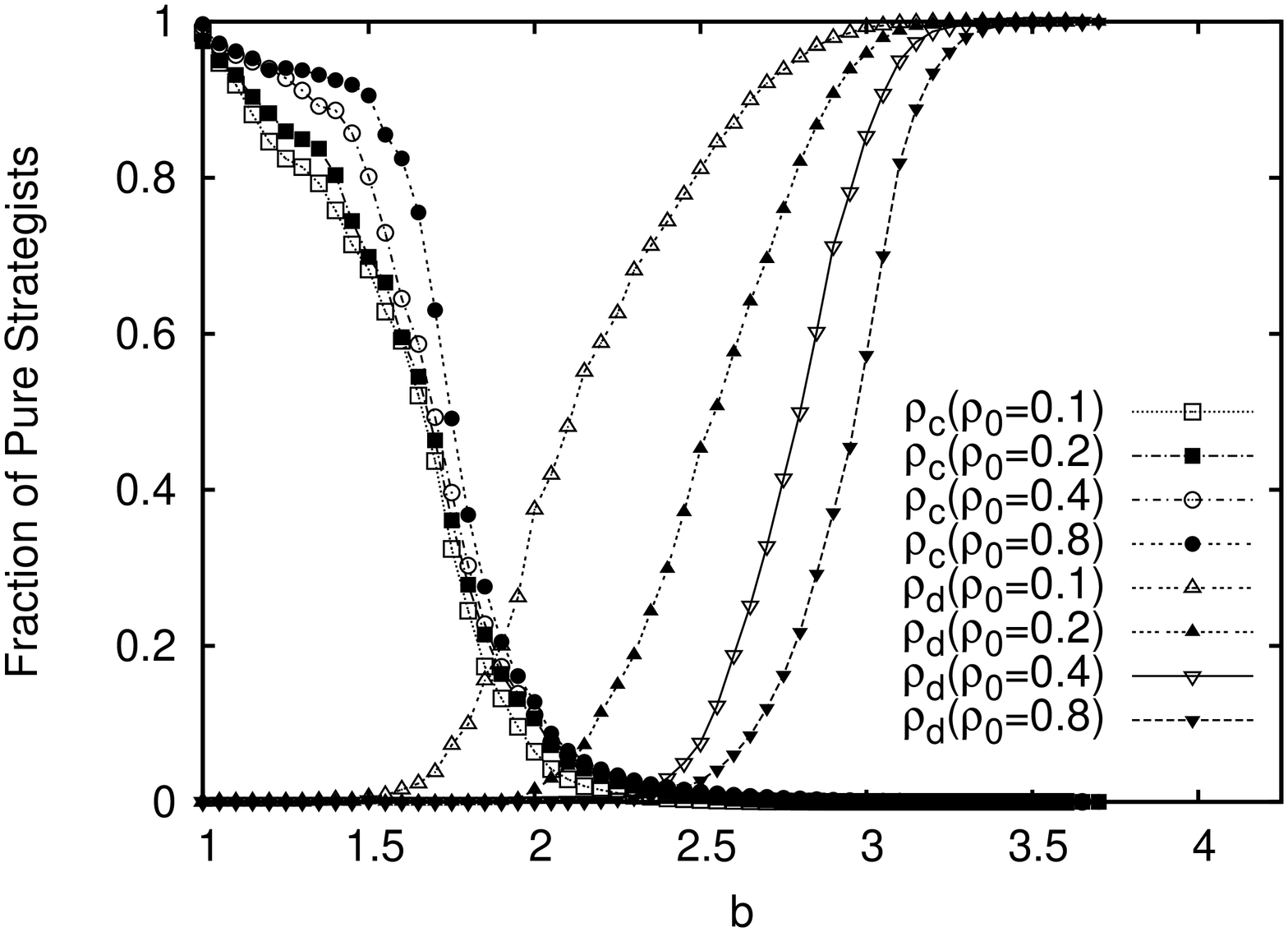,width=2.5in,angle=-90}
\end{center}
\caption{Fraction of pure strategists in ER (top) and SF (bottom) networks as
  a function of $b$ and several values of $\rho_0$. Network parameters are
  those of Fig.\ \ref{fig1}.}
\label{fig3}
\end{figure}

As stated in the introductory section \ref{intro}, it has been
reported in \cite{prlnoi} that for any asymptotic trajectory there
is a partition of the network into three sets, namely the set
${\mathcal{C}}$ of pure cooperator nodes, the set ${\mathcal{D}}$
of pure defector nodes, and the set ${\mathcal{F}}$ of fluctuating
nodes. From now on we denote by $\rho_c= \langle
\mu({\mathcal{C}}) \rangle$ the measure (relative size) of the set
of pure cooperators (averaged over initial conditions and network
realizations), and by $\rho_d= \langle \mu({\mathcal{D}}) \rangle$
that of the set of pure defectors. The behavior of $\rho_c$ and
$\rho_d$ versus the game parameter $b$ is plotted in Fig.\
\ref{fig3} for different initial distributions as a function of
the parameter $b$.

The first remarkable result is that in ER networks, the density of
pure cooperators does not depend on $\rho_0$ for the {\em whole}
range of $b$ values, in sharp contrast to the above mentioned
results for the tails of the average level of cooperation $\langle
c \rangle(b)$ (see Fig.\ \ref{fig1}). As anticipated in the
introduction (see Eq. \ref{tachan}), there are two additive
contributions to the average fraction $\langle c \rangle$ of
cooperators, namely the measure $\rho_c$ of the set of pure
cooperators, and the overall fraction of time $\bar{T}_c$ spent by
fluctuating nodes as cooperators, weighted by the relative size
$\rho_f = \langle \mu({\mathcal{F}}) \rangle$ of the fluctuating
set:

\begin{equation}
\langle c \rangle =  \rho_c + \rho_f \bar{T}_c \label{tachan2}
\end{equation}

Though the first contribution is, for ER networks, independent of
$\rho_0$, the second one does indeed depend on initial conditions,
as inferred from Fig.\ \ref{fig1} and the relation
$\rho_c+\rho_d+\rho_f=1$. High initial concentrations of
cooperators favor the fluctuating set ${\mathcal{F}}$ at the
expense of pure defectors, while the number of nodes where
fixation of cooperative strategy occurs remains largely
unaffected: $\rho_c$ is thus being mainly determined by the
network structural features. For example, in our simulations, for
large values of $b$ where $\rho_c$ is very small, we have observed
that the pure cooperator nodes form cycles. The fixation of
cooperation in these structures is assured if none of their
elements is linked to a fluctuating individual that, while playing
as a defector, is coupled to more than $k_c/b$ cooperators, where
$k_c$ is the number of cooperators attached to the element. The
number of such structures is finite in ER graphs, but as soon as
their vertices are occupied by cooperators, they will be immune to
defectors invasion.

The bottom panel of Figure.\ \ref{fig3} shows the results obtained
for BA networks. Regarding the proportion of pure cooperators, one
may differentiate two regimes: For $b<1.7$, there is a moderate
dependence of $\rho_c$ on $\rho_0$, while $\rho_c$ is almost
independent of $\rho_0$ for larger values of $b$. This behavior
correlates well with our observations \cite{pnas} on the
distribution of strategists inside the degree classes. In the
first range, pure cooperators are present in all $k$-classes and
fluctuating individuals are almost homogeneously disseminated over
low-to-intermediate $k$ classes. However, for $b>1.7$, there is a
$b$-dependent value of $k$, say $k^{*}$, such that $k$-classes are
fully occupied by pure cooperators if $k>k^{*}$ while basically no
pure cooperators are found in lower $k$-classes. In this second
range, where the degree-strategy correlations are strong, the
influence of $\rho_0$ on the asymptotic proportion of pure
cooperators is very small.

While as discussed in previous paragraphs, the proportion of pure
cooperators is either independent (ER) or slightly dependent (BA)
on initial concentration $\rho_0$, the measures of the other sets
in the partition, ${\mathcal{F}}$ and ${\mathcal{D}}$, are indeed
more influenced by the initial conditions. The dependence of
$\rho_d$ on $\rho_0$ for BA and ER networks is qualitatively the
same, that is, the proportion of pure defectors is favored (at the
expense of the fluctuating set) by a higher initial proportion of
defectors. This is consistent with the lack of degree preference
(correlation) of pure defectors, which cannot take distinctive
advantage of degree inhomogeneity: The higher their instantaneous
payoff, the more likely they invade neighboring nodes, which has
the effect of diminishing their future payoff.

\begin{figure}
\begin{center}
\epsfig{file=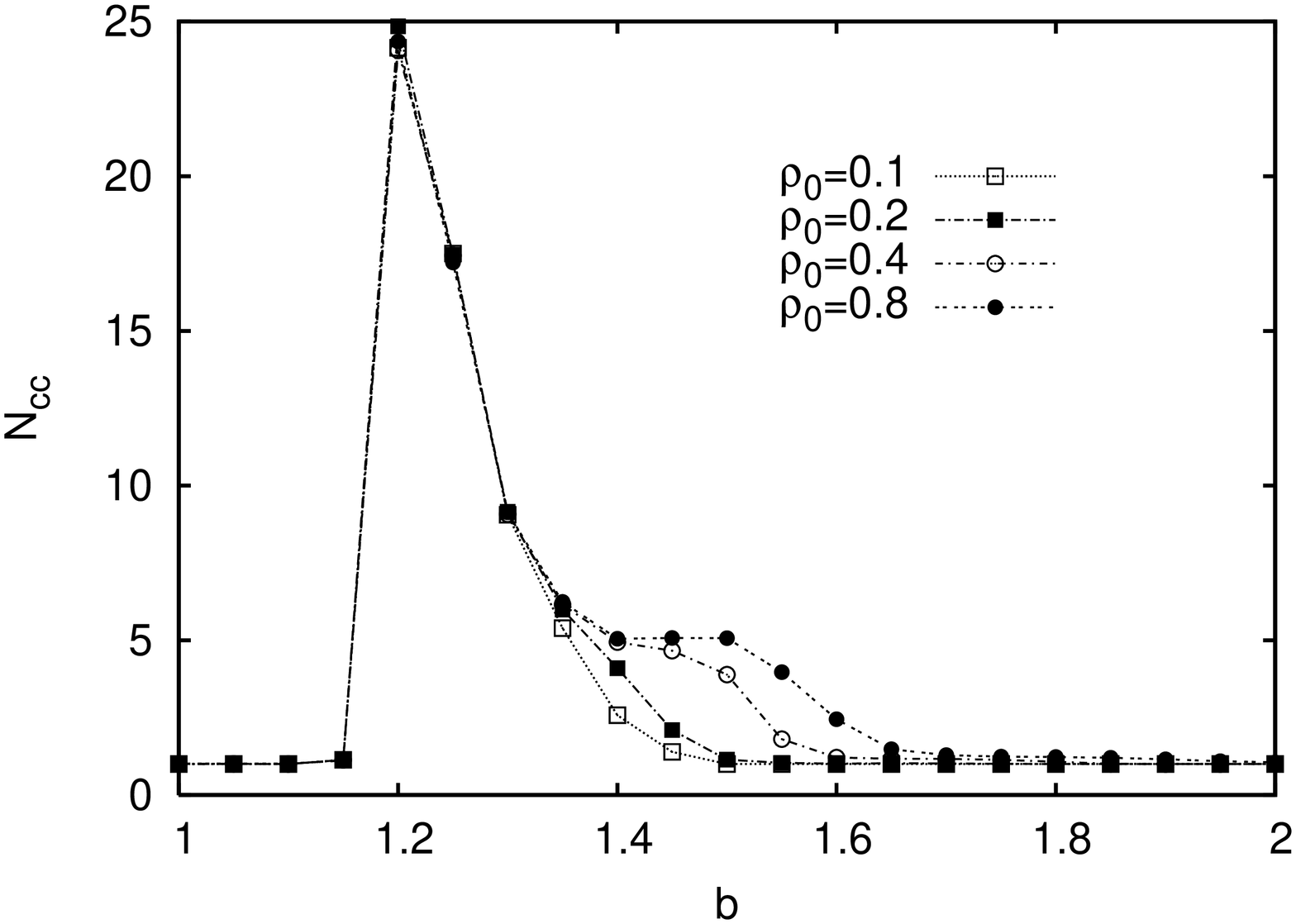,width=2.05in,angle=-90}
\epsfig{file=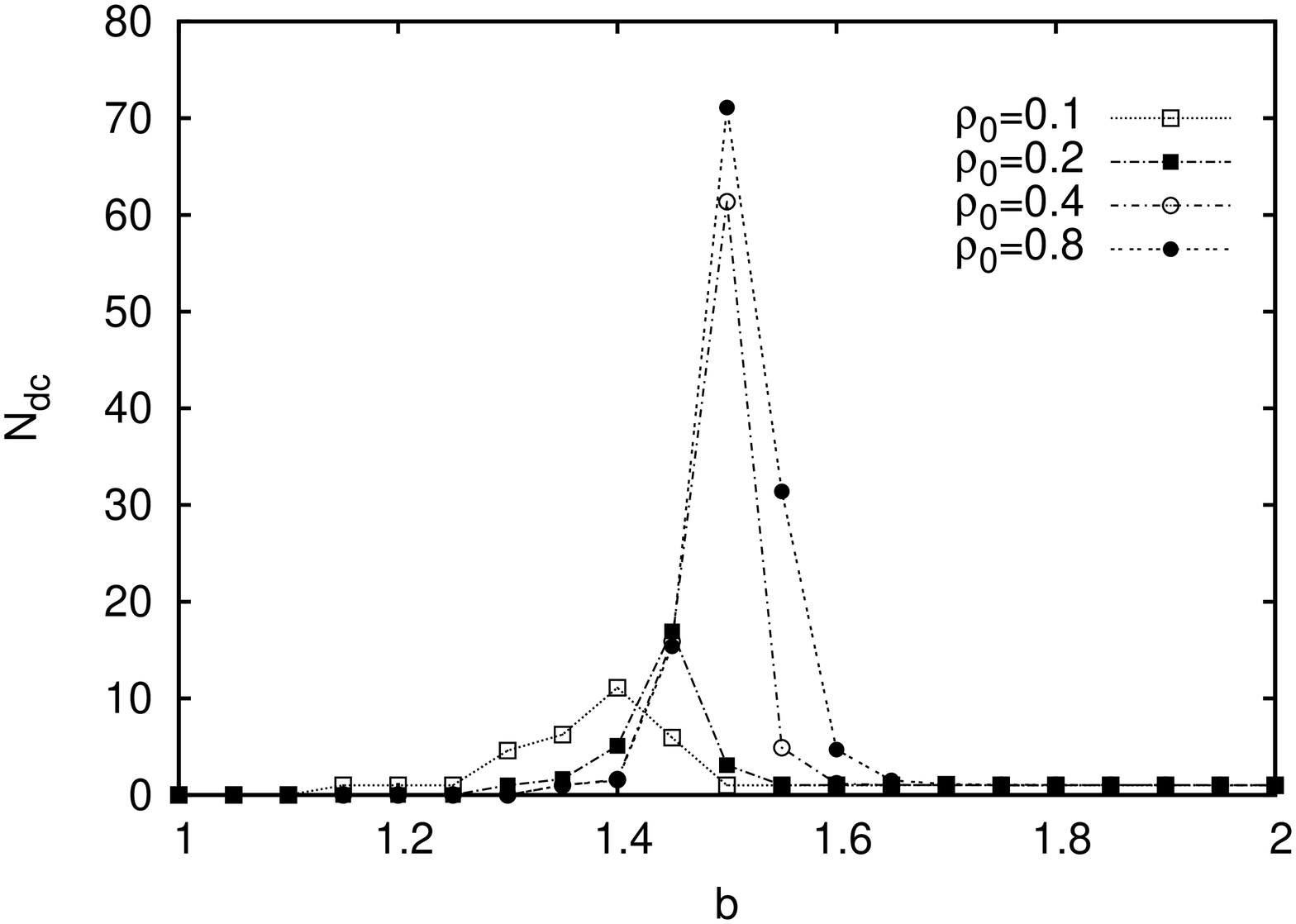,width=2.05in,angle=-90}
\epsfig{file=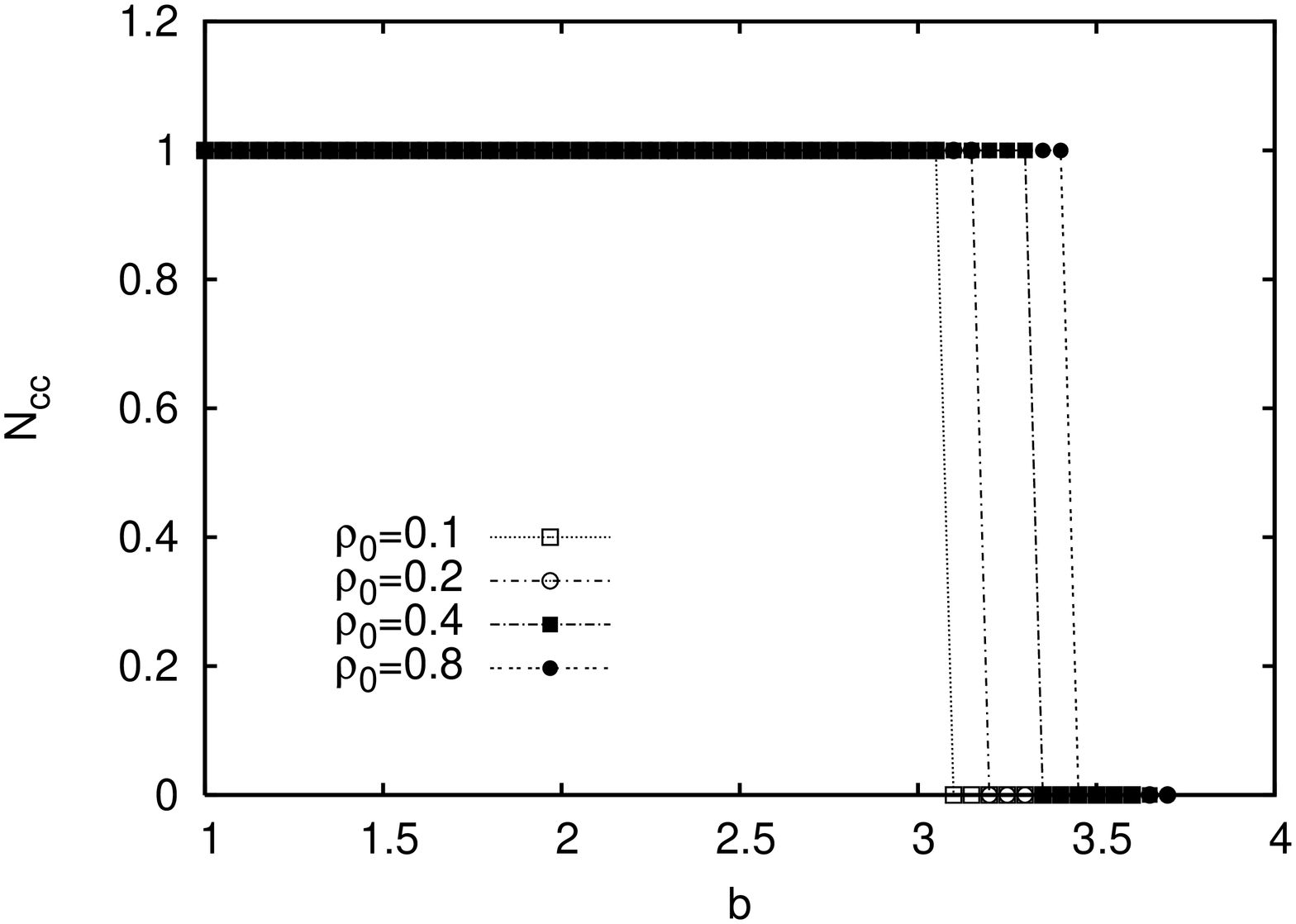,width=2.05in,angle=-90}
\epsfig{file=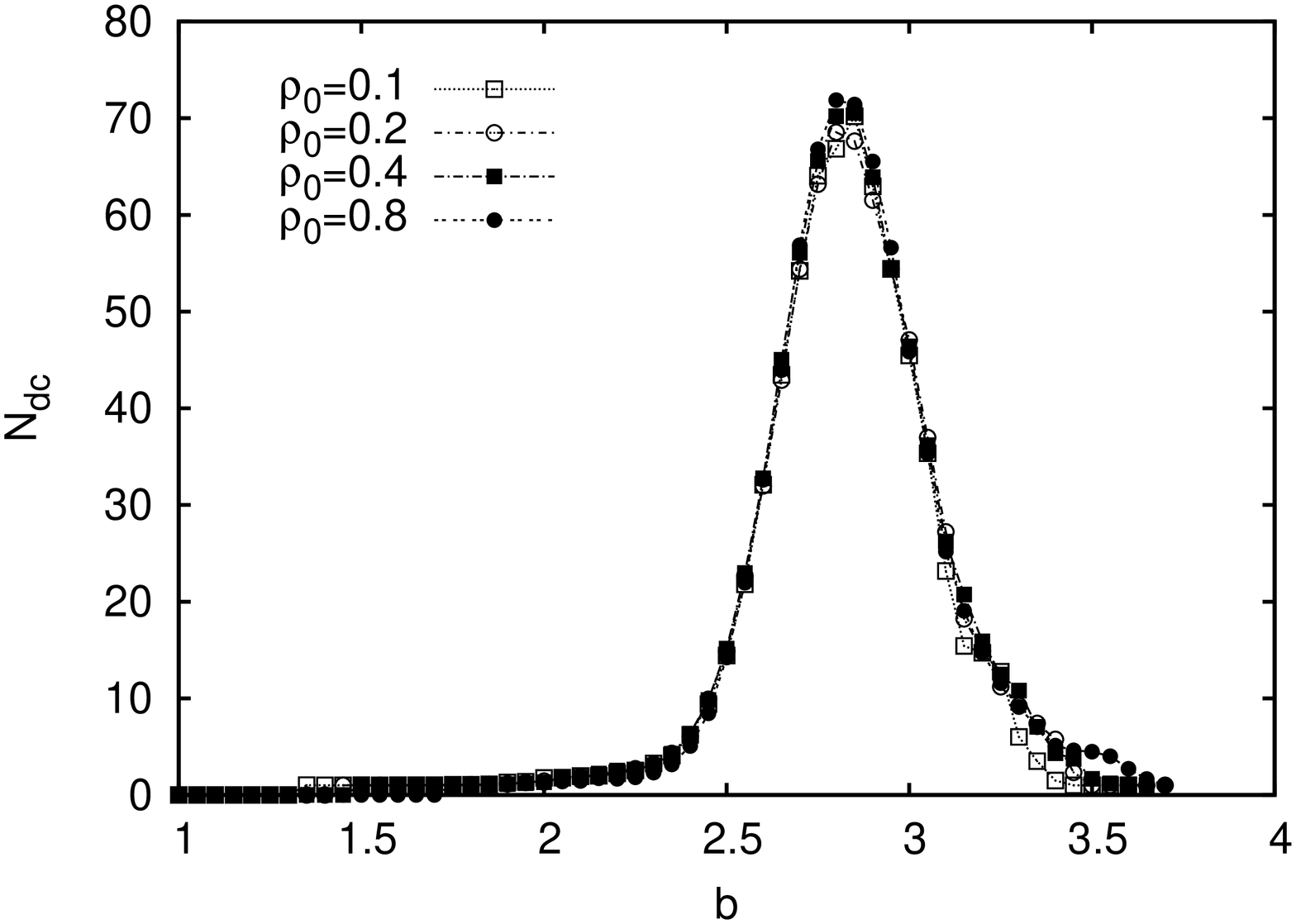,width=2.05in,angle=-90}
\end{center}
\caption{Dependence with $b$ of the number of cooperator ($N_{cc}$)
  and defector ($N_{dc}$) cores in ER graphs (top) and BA (bottom)
  networks for different values of $\rho_0$.} \label{fig4}
\end{figure}

Finally, we analyze the connectedness of the pure strategists sets, as
measured by the number of cooperator cores $N_{cc}$, and defector
cores $N_{dc}$. For BA networks, and $\rho_0=1/2$, we had reported in
\cite{prlnoi} the result that for all values of $b$ where
${\mathcal{C}}$ is not an empty set, it is connected, {\em i.e.}
$N_{cc}=1$. This result turns out to be independent of $\rho_0$ (see
Fig.\ \ref{fig4}): There is only one cooperator core in BA networks,
which contains always the most connected nodes, for any initial
fraction of cooperators. The grouping of pure cooperators into a
single connected set ${\mathcal{C}}$ allows to keep a significant
fraction of pure cooperators isolated from contacts with fluctuating
nodes. This "Eden of cooperation" inside ${\mathcal{C}}$ provides a
safe source of benefits to the individuals in the frontier,
reinforcing the resilience to invasion of the set. Pure defectors, on
the contrary, do not benefit from grouping together, and the set
${\mathcal{D}}$ appears fragmented into several defector cores. Note
that for values of $b\simeq 1$, where the set ${\mathcal{D}}$ is
empty, $N_{dc}=0$, while for very high values of $b$ defection reaches
fixation in the whole network, so that $N_{dc}=1$. Thus, $N_{dc}(b)$
must increase first and then decrease to $1$. In Fig.\ \ref{fig4} we
show the computed $N_{dc}(b)$ curves for BA networks for several
values of $\rho_0$. It is remarkable that these curves almost
collapse, in spite of the fact that the fraction $\rho_d$ of pure
defectors does indeed depend on $\rho_0$, a numerical fact for which
we have not found a plausible explanation.

In Fig.\ \ref{fig4} we also show for ER graphs $N_{cc}(b)$ and
$N_{dc}(b)$, for different values of $\rho_0$. Regarding the
number of cooperator cores, we see that except in the small range
$1.4<b<1.6$, the different curves coincide, in fair agreement with
the independence of $\rho_c$ on initial conditions. Note that in
the small interval where they do not coincide, the fraction
$\rho_c$ of pure cooperators is below $1\%$, for all values of
$\rho_0$. On the other hand, we see that for higher initial
proportion $\rho_0$ of cooperators, the set ${\mathcal{D}}$ is
more fragmented and also that $N_{dc}$ reaches its maximal values
at higher values of $b$.

\begin{figure*}
\begin{center}
\epsfig{file=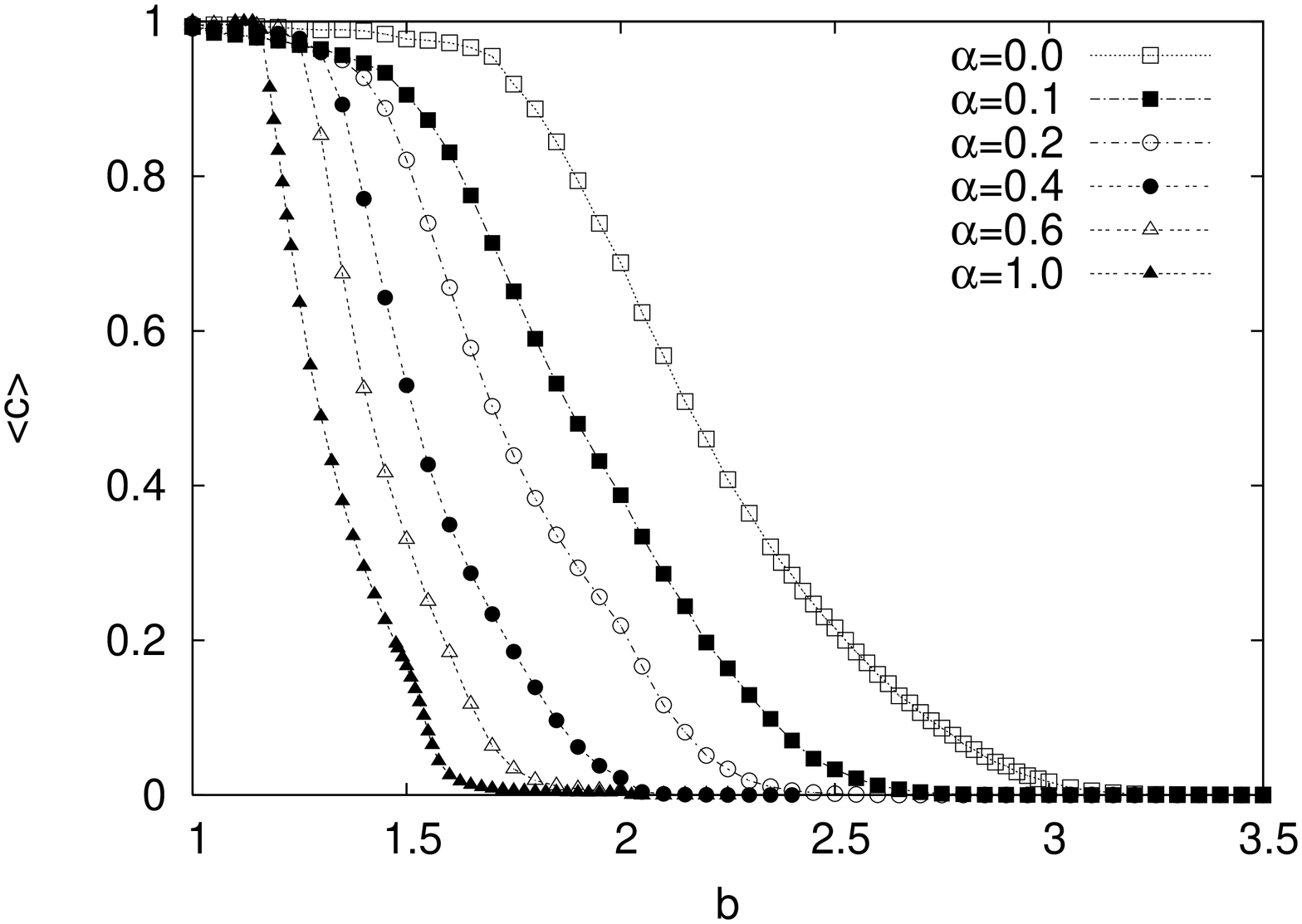,width=2.05in,angle=-90}
\epsfig{file=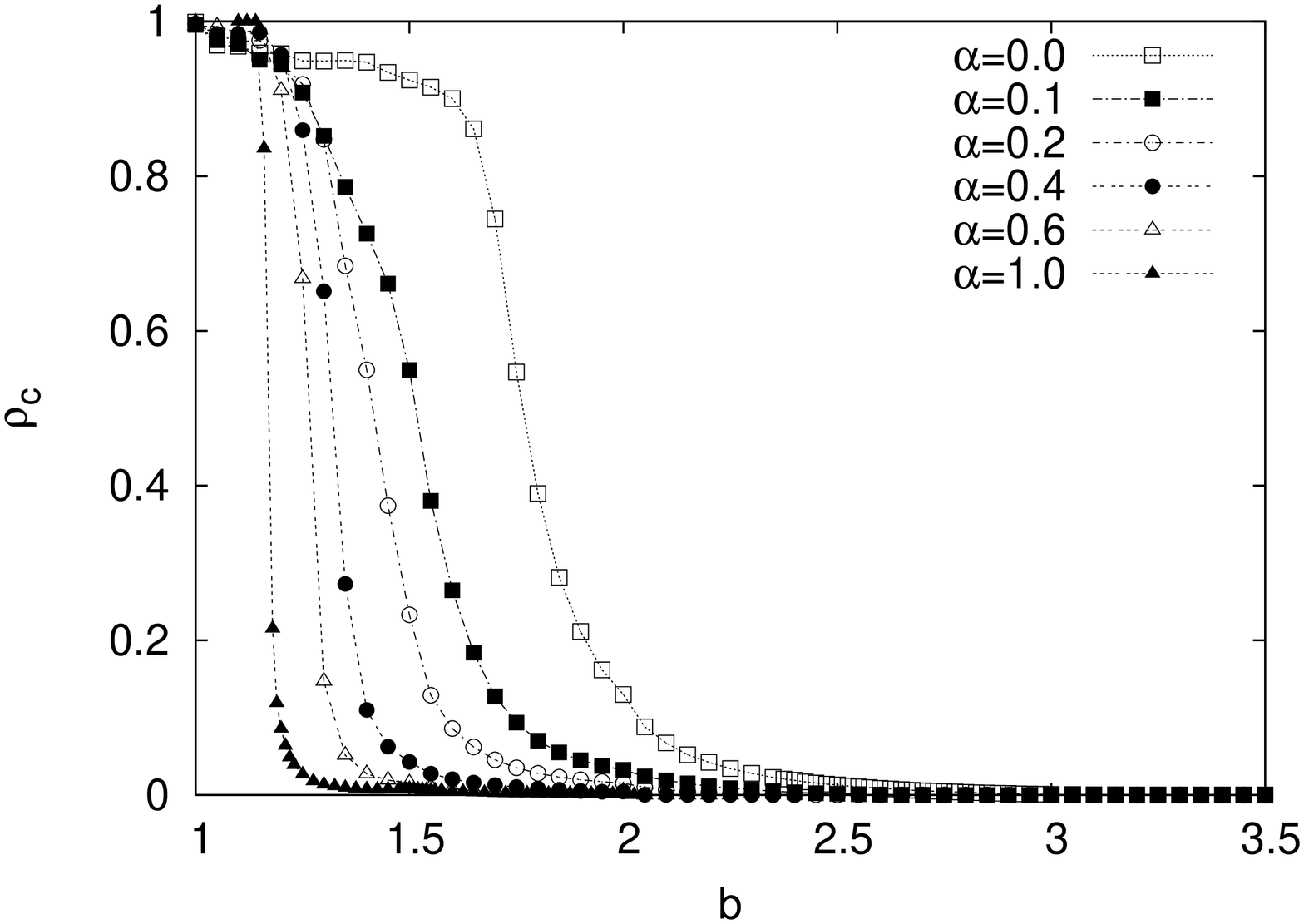,width=2.05in,angle=-90}
\epsfig{file=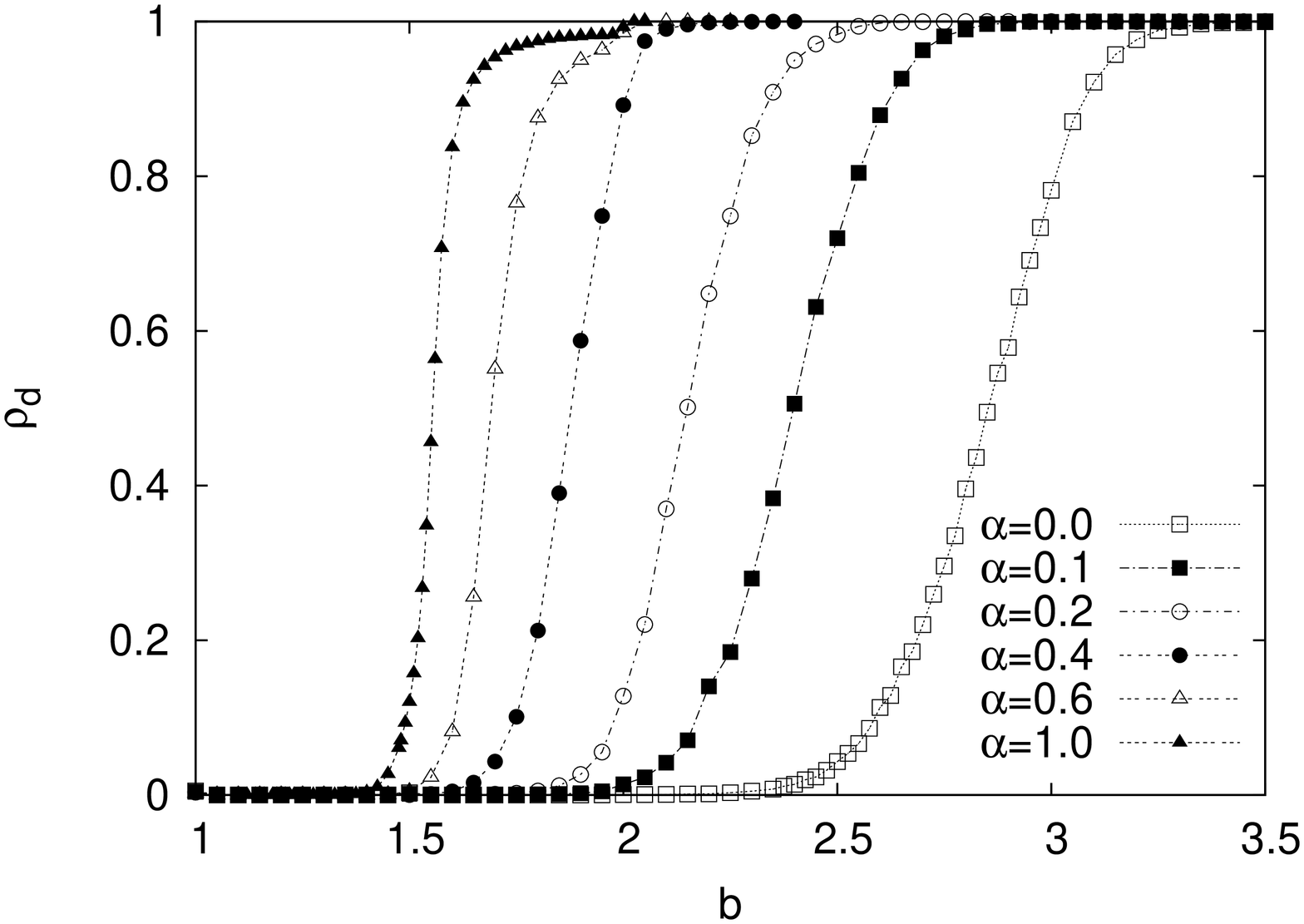,width=2.05in,angle=-90}
\end{center}
\caption{Average cooperation level and densities of strategists as
a function of $b$ for different
  values of $\alpha$. $\alpha=0$ corresponds to a BA network while $\alpha=1$
  generates an ER graph. In this case, the networks are made up of $N=2000$
  nodes and $\langle k \rangle =4$. See the main text for further details.}
\label{fig5}
\end{figure*}

\section{Influence of the degree of heterogeneity}
\label{heterogeneity}

In order to inspect how the results depend on the distribution of
nodes' degrees, we have monitored the same magnitudes studied
throughout this paper when the value of $\alpha$ varies between
$0$ and $1$. As introduced above, this makes the networks less
heterogeneous as $\alpha$ grows and approaches $1$. Figure\
\ref{fig5} shows, from left to the right, the average level of
cooperation $\langle c \rangle$, the density of pure cooperators
$\rho_c$ and the density of pure defectors $\rho_d$ as a function
of $b$ for several values of $\alpha$. In this case, the initial
distribution of cooperators was set to $\rho_0=1/2$, i.e., the
nodes have the same probability to cooperate or defect at $t=0$.
The results show that indeed the densities of pure strategists and
the average level of cooperation do depend on $\alpha$, that is to
say, the figure confirms the role played by the underlying
topology. The more homogeneous the graph is, the smaller the level
of cooperation in the system. Moreover, the transition for
different values of $\alpha$ is smooth and does not exhibit an
abrupt crossover from one kind of behavior ($\alpha=0$) to the
other ($\alpha=1$).

\begin{figure}[!tb]
\begin{center}
\epsfig{file=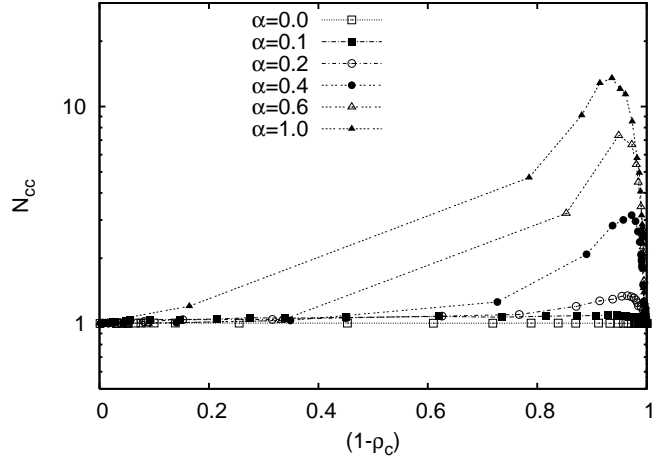,width=2.5in,angle=-90}
\end{center}
\caption{Number of cooperator cores for different networks defined by the
  value of $\alpha$ as a function of the density of nodes that are not pure
  cooperators $1-\rho_c$. Network parameters are those used in Fig.\
  \ref{fig5}.}
\label{fig6}
\end{figure}

We have also explored how nodes where strategies have reached
fixation are organized into clusters of cooperation and defection
as a function of $\alpha$. Figure\ \ref{fig6} summarizes our
computations for the number of cooperator cores. In this case, we
have represented $N_{cc}$ as a function of $1-\rho_c$ (that grows
with $b$) in order to have the same scale for different values of
$\alpha$ until cooperation breaks down. The observed dependence
with $\alpha$ is again smooth and no abrupt change in the behavior
of this magnitude occurs. It is worth stressing that as soon as
the underlying network departs from the limit $\alpha=0$
corresponding to a BA scale-free network (whose $P(k)\sim
k^{-3}$), the number of $CC$ slightly differs from $1$. This means
that some realizations give rise to more than one cluster of $CC$.
The probability to have such realizations is very small, but in
principle, they are possible. As $\alpha$ is further increased
beyond zero, it is clear that pure cooperators do not organize
anymore into a single cooperator core. We think that this
deviation is due to the fact that when $\alpha>0$ the exponent
$\gamma$ of the underlying network, which still is a scale-free
degree distribution, is larger that $3$. It is known that this
value of $\gamma$ marks the frontier of two different behaviors
when dynamical processes are run on top of complex heterogeneous
networks \cite{pv01,mpv02}. This is the case, for instance, of
epidemic spreading. For $2 <\gamma \le 3$, the second moment of
the degree distribution $P(k)$ diverges in the thermodynamic
limit, while it is finite if $\gamma>3$. As the critical
properties of the system are determined by the ratio between the
first (that remains finite for $\gamma>2$) and the second moment,
the divergence of the latter when $N\rightarrow\infty$ and $2
<\gamma \le 3$, makes the epidemic threshold null. On the
contrary, when the process takes place in networks whose
$\gamma>3$, the epidemic threshold is recovered, although no
singular behavior is associated to the critical point
\cite{pv01,mpv02}. We expect that a similar phenomenology is
behind the results shown in Fig.\ \ref{fig6}. It would be very
interesting to test this hypothesis by simulating the PD
implemented here on top of scale-free networks with an exponent in
between $2$ and $3$. As a byproduct, such a study may guide our
search of analytical insights and provides a deeper understanding
of what drives the structural organization of cooperation at the
microscopic level.

\section{Conclusions}
\label{conclusions}

Scale-free-structured populations offer to the cooperative
strategy the opportunity of positive feedback evolutionary
mechanisms making cooperation a most fitted overall strategy, in
spite of not being a best reply to itself in one-time step. We
have shown here that the enhancement of cooperation due to the
heterogeneity of the structure of connections among agents is
robust against variation of initial conditions (initial
concentration $\rho_0$ of cooperators): While both the measure of
the set ${\mathcal{C}}$ where cooperation reaches fixation, and
its connectedness properties are either independent or only
slightly dependent on $\rho_0$, the measure of the fluctuating set
${\mathcal{F}}$ and the set ${\mathcal{D}}$ where defection is
fixed, both show a clear dependence on initial conditions, for
defection cannot profit from degree heterogeneity. On the other
hand, the characteristics of the asymptotic evolutionary states of
the Prisoner's Dilemma analyzed here, show a smooth variation when
the heterogeneity of the network of interconnections is
one-parametric tuned from Poissonian to scale-free, demonstrating
a strong correlation between heterogeneity and cooperation
enhancement.

Though the numerical results presented here correspond to network
sizes $N=4000$ (in section \ref{initial}) and $N=2000$ (section
\ref{heterogeneity}), we have study also larger networks (up to
$N=10^{4}$), with no qualitative differences in the results. The
increase of network size, while keeping constant the average
degree $\langle k \rangle$, turns out to be beneficial for
cooperation, due to the fact that it has the effect of increasing
the maximal degree, and thus the range of degree values. This
further confirms how efficiently cooperation takes advantage from
degree heterogeneity.

The robustness of these results against game parameters variation
will be analyzed elsewhere \cite{pnas}, one should expect that the
network partition (${\mathcal{C}}$, ${\mathcal{D}}$,
${\mathcal{F}}$) along asymptotics stochastic trajectories is
generic in evolutionary game dynamics in graphs, for the kind of
stochastic updating rule considered here. Our results also suggest
that more works are needed in order to fully characterize the
behavior of the PD game in heterogeneous graphs. The use of real
networks, with emphasis on the role of mesoscopic (community)
structures is addressed in \cite{lozano}. Of particular interest
would be to perform the sort of analysis carried out here in
scale-free networks with an exponent $2 < \gamma <3$, which will
make it feasible to connect evolutionary dynamics with other
dynamical processes taking place on top of scale-free networks.
Our hope is that this sort of study might provide a deeper
understanding of what is going on at the microscopic level and
might help to comprehend what universal mechanisms drive the
evolution of complex heterogeneous networks as well as the reasons
behind their ubiquitous presence in nature.

\begin{acknowledgments}
  We thank A. Arenas, J.\ M.\ Pacheco and A. S\'{a}nchez for helpful
  comments and discussions.  J.G.G. and Y.M. are supported by MEC
  through a FPU grant and the Ram\'{o}n y Cajal Program,
  respectively. This work has been partially supported by the Spanish
  DGICYT Projects FIS2006-12781-C02-01, and FIS2005-00337.
\end{acknowledgments}

\end{document}